\documentclass[fleqn,aps,pra,amsmath,amssymb,reprint,showpacs]{revtex4-1}

\usepackage{amsmath}
\usepackage{graphicx}
\usepackage{dcolumn}
\usepackage{bm}
\usepackage{textcomp}

\begin{document}
\title{Accessing the $\rm 5S_{1/2} \rightarrow 5D_{5/2}$ two-photon transition in Rb using a diode laser system}
 \author{Ketan D. Rathod}
 \affiliation{Department of Physics, Indian Institute of
 Science, Bangalore 560\,012, India}
 \author{Vasant Natarajan}
 \affiliation{Department of Physics, Indian Institute of
 Science, Bangalore 560\,012, India}
 \email{vasant@physics.iisc.ernet.in}
 \homepage{www.physics.iisc.ernet.in/~vasant}

\begin{abstract}
We report observation of the $\rm 5S_{1/2} \rightarrow 5D_{5/2}$ two-photon transition in Rb vapor at 778 nm, using an external cavity diode laser system and a heated vapor cell. The spectra in the two isotopes show well-resolved hyperfine transitions. The peaks are Doppler free, and have a Lorentzian lineshape with a typical linewidth of 2.2 MHz. This linewidth is larger than the natural linewidth of 300 kHz, but is still 5--10 times smaller than the linewidth for single-photon transitions in the D$_2$ line. Since the absolute frequency of this transition is measured with 8 kHz precision, it can form a better secondary reference in the optical regime compared to the D$_2$ line.
\end{abstract}



\maketitle

\section{Introduction}
Precise measurement of \textit{optical} frequencies (of order 10$^{15}$ Hz) is complicated by the fact that the SI unit of time is defined in terms of a microwave hyperfine transition in $^{133}$Cs---one at $9.19 \times 10^9$ Hz. Thus the frequency measurement requires a span of more than 5 orders-of-magnitude with no phase error, not an easy task. In recent times, the use of a femtosecond frequency comb has enabled such a precise span \cite{HAL06}. An alternative technique, which is lower cost and simpler, is to use an optical transition as a secondary frequency standard, and compare an unknown frequency with the reference. In our laboratory, we have pioneered such a technique where we use a ring-cavity resonator as a transfer cavity between the unknown and reference frequencies \cite{BDN03,BDN04,DBB06,SMM12}. The reference in these experiments was a laser locked to a hyperfine transition in the D$_2$ line of $^{87}$Rb (at 780 nm), whose absolute frequency has been measured using a frequency chain to 10 kHz precision \cite{YSJ96}. 

The technique is particularly suited to the measurement of electric-dipole allowed (E1) transitions because the $Q$ for such transitions is about $10^8$. Thus a precision of 10 kHz (which corresponds to a relative uncertainty of $10^{-11}$ in the measurement of an optical transition) requires splitting the spectral line by 1 in 1000, which is a considerable experimental challenge. In other words, the precision of measurements on such transitions is limited by the linewidth of the transition itself and not the measurement technique, which makes the ring-cavity resonator a competitive alternative to the comb method.

Since the natural linewidth of the D$_2$ line used as the frequency reference is about 6 MHz, reaching a precision of 10 kHz in the lock point itself requires better than 1 in 500 reproducibility. Therefore an attractive option is to use a two-photon transition, which are forbidden from E1 selection rules, and consequently have sub-MHz natural linewidth. One such transition is the 5S$_{5/2} \rightarrow $ 5D$_{5/2}$ transition in Rb with a natural linewidth of 300 kHz. The absolute frequency of this transition (at 778 nm) has been measured by Nez et al.\ with 8 kHz precision \cite{NBF93}, thus making it a useful secondary reference. In addition, a two-photon transition can be made Doppler free if the two photons come from counter-propagating beams, because only zero-velocity atoms are resonant with both beams.

In this work, we report observation of the 5S$_{1/2}\rightarrow $ 5D$_{5/2}$ two-photon transition in Rb using a diode laser system. We are able to resolve the various hyperfine transitions in the two isotopes, $^{85}$Rb and $^{87}$Rb.  We obtain a linewidth of about 2.2 MHz (limited by optical feedback into the diode laser system), which is about 5--10 times better than the linewidth obtained for the D$_2$ line used as a frequency reference in the ring-cavity resonator technique.

\section{Experimental details}

The diode laser system is a home-built grating stabilized diode laser \cite{BRW01}. The laser has a free running wave length of 784 nm and a maximum power of 120 mW, which reduces to 70 mW after feedback. The linewidth after feedback is about 1 MHz. Using a combination of operating temperature and grating angle, the system is brought to operate near 778 nm required for accessing the two-photon transition. Coarse tuning is achieved using a commercial wavemeter (WS Ultimate from High Finesse Laser \& Electronic Systems). The laser beam has elliptic cross-section with $1/e^2 $ diameter of 3 mm $\times$ 5 mm.  It is linearly polarized.

\begin{figure}
\centering
\resizebox{0.95\columnwidth}{!}{\includegraphics{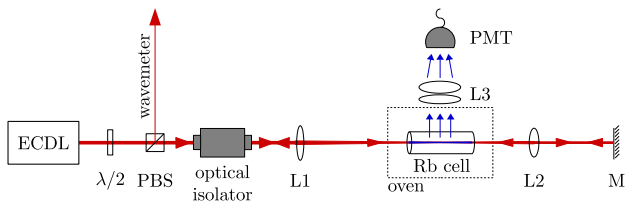}}
\caption{(Color online) Experimental schematic for two-photon spectroscopy in Rb. The vapor cell is heated to a temperature of $135^\circ$C. Figure key --- ECDL: external cavity diode laser; $\lambda$/2: half-wave plate; PBS: Polarization beam splitter cube; L1,L2,L3: lenses; M: mirror; PMT: photo multiplier tube.}
\label{twophotonsetup}
\end{figure}

The schematic of the experiment is shown in the Fig.\ \ref{twophotonsetup}. The laser output first passes through a two-stage optical isolator (with 60 dB isolation), and then through a Rb vapor cell heated to $135^\circ $C. The cell is cylindrical in shape with 25 mm diameter and 50 mm length. It is pure (no buffer gas) and contains the two isotopes of Rb in their natural abundances --- $ 72\% \ ^{85}$Rb and $28\% \ ^{87}$Rb. The laser beam is focused into the cell using a biconvex lens (L1) of focal length 300 mm. After the cell, the beam passes through a second lens (L2) of the focal length 100 mm before being retro-reflected from a mirror.

\begin{figure}
\centering
\resizebox{0.95\columnwidth}{!}{\includegraphics{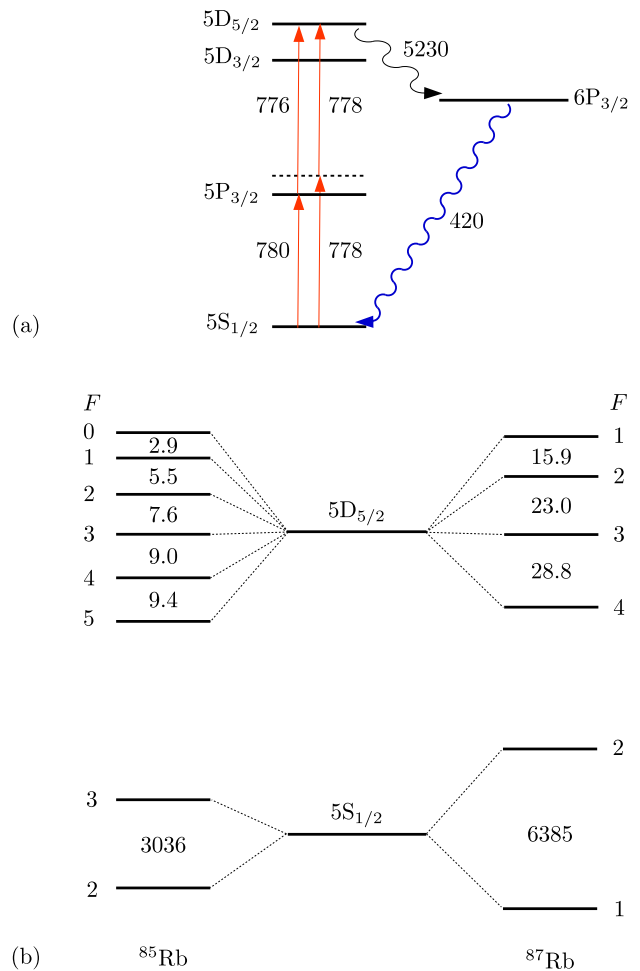}}
\caption{(Color online) (a) Energy levels of Rb. The laser at $778$ nm drives the two-photon transition through an intermediate virtual state (dotted line). The two single-photon transitions at $780$ nm ($\rm D_2$ line) and at $776$ nm are also shown. Population in the 5D state decays to the ground state by radiative cascade through the intermediate 6P state---the resulting blue fluorescence at $420$ nm is the detected signal. (b) Hyperfine levels in the two isotopes in the $\rm 5S_{1/2}$ (ground) and $\rm 5D_{5/2}$ states. All intervals are in MHz. Note that the hyperfine structure in the two states are opposite.}
\label{twophotonenergylevel}
\end{figure}

The relevant energy levels in the two isotopes are shown in Fig.\ \ref{twophotonenergylevel}. The isotope $^{85}$Rb has $I=5/2$ and therefore 6 
hyperfine levels in the D$_{5/2}$ state. The other isotope $^{87}$Rb has $I=3/2$ and therefore 4 hyperfine levels in the D$_{5/2}$ state. The signal from population in the D$_{5/2}$ state is the blue fluorescence at 420 nm due to radiative cascade through the intermediate 6P$_{3/2}$ state, as shown in Fig.\ \ref{twophotonenergylevel}(a).

The fluorescence signal (which is also visible to the naked eye) is measured by a R928 Hamamatsu photomultiplier tube (PMT) using a pair of lenses (L3 in the schematic figure). The solid angle subtended by the lens system is such that about $5\%$ of the fluorescence light is collected. The PMT has a 420 nm interference filter in front with peak transmission of $72\%$.

\section{Results and Discussion}

\begin{figure}
\centering
\resizebox{0.95\columnwidth}{!}{\includegraphics{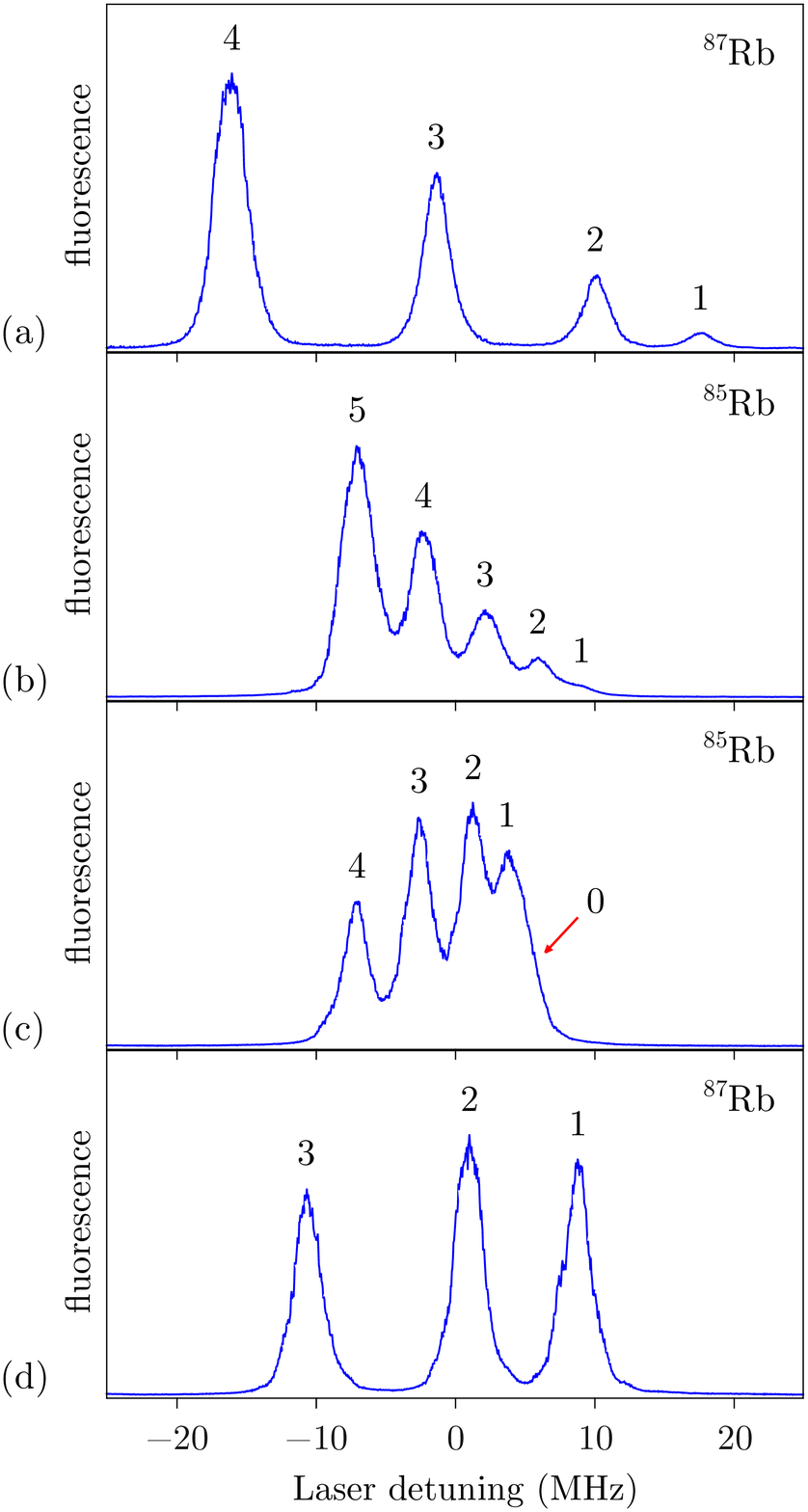}}
\caption{(Color online) The $\rm 5S_{1/2}\rightarrow 5D_{5/2}$ two-photon transitions in the two isotopes of Rb. The selection rule is $\Delta F \leq 2$. (a) $^{87}$Rb: $F=2 \rightarrow F'=4$ to $1$. (b) $^{85}$Rb:  $F=3\rightarrow F'=5$ to $1$. (c) $^{85}$Rb:  $F=2\rightarrow F'=4$ to $0$; the $F'=1$ and $0$ levels are not resolved---the location of $F'=0$ is indicated by an arrow. (d) $^{87}$Rb: $F=1\rightarrow F'=3$ to $1$.}
\label{twophotonsignals}
\end{figure}

The measured two-photon spectra in the two isotopes are shown in Fig.\ \ref{twophotonsignals}. There are four spectra corresponding to the two ground hyperfine levels in each isotope---$F=2,3$ in $^{85}$Rb and $F=1,2$ in $^{85}$Rb. The selection rules for a two-photon transition are $\Delta F \leq 2 $, therefore each spectrum has several well-resolved peaks corresponding to this rule. Only the $F=2 \rightarrow 1,0$ peaks in $^{85}$Rb are not resolved. As expected, the spectra appear on a flat Doppler-free background. The resolved peaks have a Lorentzian lineshape with a linewidth of about 2.2 MHz. This increase from the natural linewidth of 300 kHz is due to optical feedback into the laser from the system. Despite using an  optical isolator in front of the laser, stable spectra of the kind shown are obtained only when the counter-propagating beam is slightly misaligned. For comparison, the linewidth seen in Ref.\ \cite{NBF93} was 500 kHz, but the experiment was done with a frequency stabilized Ti-sapphire laser with frequency uncertainty of only 2 kHz, and a laser that was much less sensitive to optical feedback.

\begin{figure}[h]
\centering
\resizebox{0.95\columnwidth}{!}{\includegraphics{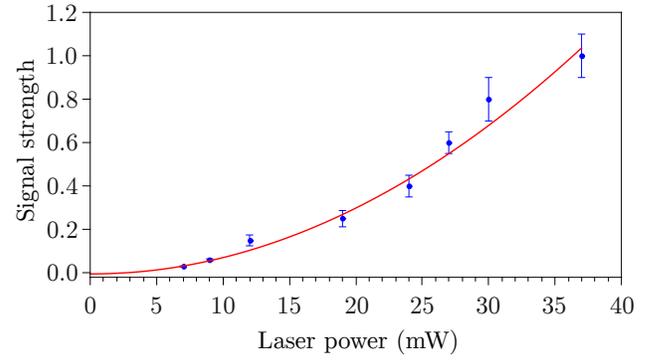}}
\caption{(Color online) Fluorescence amplitude for the two-photon transitions as a function of
	laser power. The error bars show the fluctuation observed. The fluctuations are larger
	at higher powers due to increased optical feedback.}
\label{twophotonsquarelaw}
\end{figure}

To verify that it is a two-photon transition, we have studied the signal strength as function of laser power. For a two-photon transition, the rate varies as the square of the laser power. The results are shown in Fig.\ \ref{twophotonsquarelaw}, along with a quadratic fit to the data. The error bars represent the standard deviation for each data set, which consists of 20 points. The quadratic curve fits the data well within the error bars, highlighting the two-photon nature of the spectrum.

\clearpage

\section{Conclusion}
In summary, we have observed the $\rm 5S_{1/2}\rightarrow 5D_{5/2}$ two-photon transitions in a Rb vapor cell using a diode laser system. The various hyperfine transitions in the two isotopes are well resolved and have a linewidth of about 2.2 MHz. This is somewhat larger than the 300 kHz natural linewidth, and is due to laser jitter caused by optical feedback. But the linewidth is still about an order-of-magnitude smaller than the linewidth obtained for the D$_2$ line, which is used in our laboratory as a frequency reference in a ring-cavity resonator \cite{SMM12}. This makes the two-photon transition a useful frequency reference for improved precision in the ring-cavity technique.

\begin{acknowledgments}
This work was supported by the Department of Science and Technology, India. K.D.R acknowledges financial support from Council for Scientific and Industrial Research, India. 
\end{acknowledgments}


\end{document}